\documentclass[preprint,aps]{revtex4}
\usepackage{graphicx}
\usepackage{subcaption}
\usepackage{bm}
\begin{document}
	
\title  {Semileptonic $B_c$ meson decays to S-wave charmonium states} 
\author {Sonali Patnaik$^{1}$, Lopamudra Nayak$^{1}$, P. C. Dash$^{1}$, 
	Susmita Kar$^{2}$\footnote{email address:skar09.sk@gmail.com}, 
	N. Barik$^{3}$}

\affiliation{$^{1}$ Department of Physics,
	Siksha 'O' Anusandhan Deemed to be University, Bhubaneswar-751030, India\\
	$^{2}$ Department of Physics, North Orissa University, Baripada-757003, India\\ 
	$^{3}$ Department of Physics, Utkal University, Bhubaneswar-751004, India}

\begin{abstract}
We study the semileptonic decays of $B_c$ meson to S-wave charmonium states in the framework of relativistic independent quark model based on an average flavor-independent confining potential $U(r)$ in the scalar-vector harmonic form $U(r)=\frac{1}{2}(1+\gamma^0)(ar^2+V_0)$, where ($a$, $V_0$) are the potential parameters.The form factors for $B_c^+\to \eta_c /\psi e^+\nu_e$ transitions are studied in the physical kinematic range.
Our predicted branching ratios (BR) for transitions to ground state charmonia are found comparatively large $\sim $ $10^{-2}$, compared to those for transitions to radially excited 2S and 3S states. Like all other mpdel predictions, our predicted BR are obtained in the hierarchy: BR($B_c^+\to \eta_c /\psi (3S)$) $<$ BR($B_c^+\to \eta_c/ \psi (2S)$) $<$ BR($B_c^+\to \eta_c /\psi (1S)$). The longitudinal ($\Gamma_L$) and transverse polarization ($\Gamma_T$) for $B_c \to \psi(ns)$ decay modes are predicted in the small and large $q^2$ - region as well as in the whole physical region. The ratios for such transitions are obtained $\frac {\Gamma_L}{\Gamma_T} < 1$  throughout the kinematic range which means the $B_c^+$ meson transitions to vector meson charmonium states take place predominantly in transverse polarization mode. The theoretical predictions on these transitions could be tested in the on-going and forthcoming experiments at LHCb. 
	
\end{abstract}
\maketitle 

\section{Introduction}
Ever since the discovery of $B_c$ meson in Fermilab by the Collider Detector (CDF) Collaborations \cite{A1} in 1998, the experimental probe to detect its family members in their ground and excited states continues over last two decades. With the observation of $B_c$ meson at hadron collider at Tevatron \cite{A2,A3}, a detailed study of $B_c$ family members is expected at the LHC, where the available energy is more and luminosity is much higher. The lifetime of $B_c$ has been measured \cite{A4,A5,A6,A7} using decay channels: ${B_c}^{\pm} \rightarrow J/\psi e^{\pm} \nu_e$ and ${B_c}^{\pm} \rightarrow J/\psi \pi^ {\pm}$.  A more precise measurement of $B_c$-lifetime: $\tau_{B_c}=0.51^{+0.18}_{-0.16}(stat.)\pm 0.03(syst.)\;ps$ and its mass: $M =6.40\pm 0.39\pm 0.13\;GeV$ have been obtained \cite{A8} using the decay mode $B_c \rightarrow J/\psi \mu \nu_\mu X$, where X denotes any possible additional particle in the final state. The branching fraction for $B_c^+ \rightarrow J/\psi\pi^+ $ relative to that of $B_c^+ \rightarrow J/\psi\mu^+ \nu_{\mu}$ has been measured by LHCb Collaborations yielding \cite{A9}:

\begin{center}
	$\frac {BR (B_c\to J/\psi \pi^+)}{BR (B_c\to J/\psi \mu^+ \nu_{\mu})} = 0.0469 \pm 0.0028(stat)\pm 0.0046(syst)$
\end{center}

Recently, ATLAS Collaboration at LHC has detected excited $B_c$ state \cite {A10} through the channel $B_c^{\pm}(2S) \rightarrow B_c^{\pm}(1S)\pi^+\pi^- $ by using 4.9 $fb^{-1}$ of 7 TeV and 19.2 $fb^{-1}$ of 8 TeV pp collision data which yielded $B_c$(2S) meson mass $\sim$ $6842\pm 4\pm 5$ MeV. Although masses of the ground and first excited state of $B_c$ with $J^P$ = $0^-$ have been measured, it has not yet been possible to detect its higher excited states and even the ground state of $B_c^*$. Hopefully with the available energy and higher luminosity at LHC and at $Z_0$ factory, the event accumulation rate for these undetected states can be enhanced in near future providing scope for detailed studies of $B_c$ and $B_c^*$ counterpart.The recent observed data and possibility of high statistics $B_c$ events expected in upcoming experiments provide necessary motivation to investigate the semileptonic $B_c$ meson decays to charmonium states which are easier to identify in the experiment.\\

The $B_c$ meson has aroused a great deal of theoretical interest due to its outstanding features. It is the lowest bound state of two heavy constituent quarks (charm and bottom) with open (explicit) flavor unlike the symmetric heavy quarkonium ($b\bar b, c\bar c$) states. The charmonium ($c\bar c$) and bottomonium ($b\bar b$) with hidden flavors decay via strong and electromagnetic interactions whereas $B_c$ meson with open flavors decay only via weak interaction since it lies below the B$\bar D$ threshold. Therefore it has comparatively long lifetime and very rich weak decay channels with sizable branching ratios. Thus $B_c$-meson provides a unique window into heavy quark dynamics and give scope for independent test of quantum chromodynamics.  The study of semileptonic decays, in particular, is significant because it not only helps in extracting accurate values of the Cabbibo-Kobayashi-Masakawa (CKM) matrix element but also helps in separating the effect of strong interaction from that of weak interaction into a set of lorentz invariant form factors. The analysis of semileptonic decays is therefore reduced essentially to the calculation of relevant weak form factors.\\

Semileptonic $B_c$ decays have been widely studied in the literature. Although it is not possible to cite them all, a few noteworthy ones are: potential models \cite{A11,A12,A13,A14,A15,A16,A17,A18}, non-relativistic qurak models \cite{A19,A20}, relativistic quark models \cite{A21,A22,A23,A24,A25}, instantaneous non-relativistic approach  to BS equation \cite{A26}, relativistic quark model on BS approach \cite{A27}, non-relativistic QCD \cite{A28,A20,A30}, light-cone QCD sum rule \cite{A31,A32,A33}, covariant light-front model \cite{A34}, light-front quark model constrained by the variational principle for QCD motivated effective Hamiltonian \cite{A35}, light-front quark model \cite {A36}, QCD potential model\cite { A37,A38,A39}, perturbative QCD approach \cite{A40,A41,A42,A43,A44}, constituent quark model \cite{A45,A46,A47,A48,A49} 
and Isgur, Scora, Grinstein and Wise (ISGW) model \cite{A50}. In this context one would also like to refer to review paper \cite{A51} and references there in. In this paper we would like to extend the applicability of our relativistic independent quark (RIQ) model \cite {A52,A53,A54,A55,A56,A57,A58,A59} which has already been tested in describing wide ranging hadronic phenomena including the static properties of hadrons \cite{A52} and various decays such as radiative, weak radiative, rare radiative \cite {A53}, leptonic, weak leptonic, radiative leptonic \cite {A54}, semileptonic \cite {A55,A56} and nonleptonic \cite{A57} decays of hadrons in the light and heavy flavor sector. In our previous work on semileptonic $B_c$ meson decays, we consider the participating mesons in their respective ground state only. In view of observed $B_c$(2S) states and possible detection of  higher $B_c$(nS) states  ($n>2$) as well as $B_c^*$ (1S) state at LHC and $Z_0$ factory in near future, it is worthwhile to predict energetically allowed semileptonic $B_c$ decays to excited charmonium states too. In fact a number of theoretical approaches in this direction have already appeared in the literature. Being inspired by our recent prediction of magnetic dipole \cite{A58} and electromagnetic \cite{A59} transitions of $B_c$ and $B_c^*$ mesons in their ground and possible excited states we extend our previous work \cite{A56} to analyze $B_c^+(nS)\to \eta_c(nS)/\psi(nS) e^+\nu_e$ decays, where the radial quantum numbers n=1,2,3. We don't consider here the decay modes with higher 4S charmonia as their properties are not yet understood well.\\

Here we would like to note few points that motivate us to undertake this exercise: (1) The relevant form factors representing the weak decay amplitudes are expected to have their $q^2$ dependence (where $q^2$ denotes the four momentum transfer squared) over entire kinematic range. In some of the theoretical approaches cited above the weak decay form factors and their $q^2$ dependence are determined first with an end point normalization at either $q^2$ =0 or $q^2$ = $q^2_{max}$ and then using some phenomenological monopole/dipole/gaussian ansatz they are extrapolated to the whole physical region. In order to avoid possible uncertainties in the calculation we shall not resort to any such phenomenological ansatz and instead study the $q^2$ dependence of relevant form factors in the allowed kinematic range: $0\leq\ q^2\leq\ (M-m)^2$, where M and m refer to mass of the parent and daughter meson, respectively.
(2) In our previous analysis \cite{A55,A56} two weak form factors $a_+$ and $a_-$ corresponding to $0^-\to 1^-$ semileptonic transition are found to be equal under a simplifying assumption. On closure scrutiny it is realized that, $a^+$ is not strictly equals to $a_-$. It is not necessary to invoke any kind of simplifying assumption but nonetheless get model expressions seperately for $a^+$ and $a^-$. (3) In this work we intend to predict the BR for decay channels involving the ground as well as excited charmonia in the final state and compare our results with other model predictions. (4) Finally, we shall update some input hadronic parameters according to Particle Data Group 2018 \cite{A60} in our calculation.\\ 
This paper is organized as follows: In Section-II we provide the general formalism and kinematics of $B_c$ meson semileptonic decays. Section-III briefly describes the framework of RIQ model and extraction of model expression for the weak form factors. In Section-IV we provide our numerical results and discussion. Section-V encompasses our summary and conclusion.   

\section{General Formalism and Kinematics} 
The invariant transition matrix element for exclusive semileptonic decays such as $B_c^+ \to X e^+\nu_e$ is written as \cite{A54,A55,A61}
\begin{equation}
{\cal M}={{G_F}\over {\sqrt {2}}}V_{bc}L^{\mu}H_{\mu}
\end{equation}
where, X denotes $\eta_c$ or $J/\psi$,  $G_F$ is the effective Fermi-coupling constant, $V_{bc}$ 
is the CKM parameter, $L^{\mu}$ and $H_{\mu}$, respectively are the leptonic and hadronic amplitudes expressed as:
\begin{equation}
L^\mu={\bar u}_e(\vec p_1,\delta_1)\gamma^{\mu}(1-\gamma_5)v_{\nu_e}(\vec p_2\;,
\delta_2),
\end{equation}
\begin{equation}
H_{\mu}=\;<X(\vec k, S_X)\mid J^h_{\mu}(0)\mid B_c(\vec p, S_{B_c})>
\end{equation}
Here $J^h_{\mu}=V_{\mu}-A_{\mu}$ is the vector-axial vector current. We take here ($M,m$) to be the mass, (p, k) the four momentum and ($S_{B_c}, S_{X}$) the
spin projection of parent and daughter meson, respectively.
$q=(p-k)=(p_1+p_2)$ is four momentum transfer where ($p_1, p_2$) are four momenta of the lepton pair.
\begin{figure}
	\begin{center}
		\includegraphics[width=15cm,height=8cm]{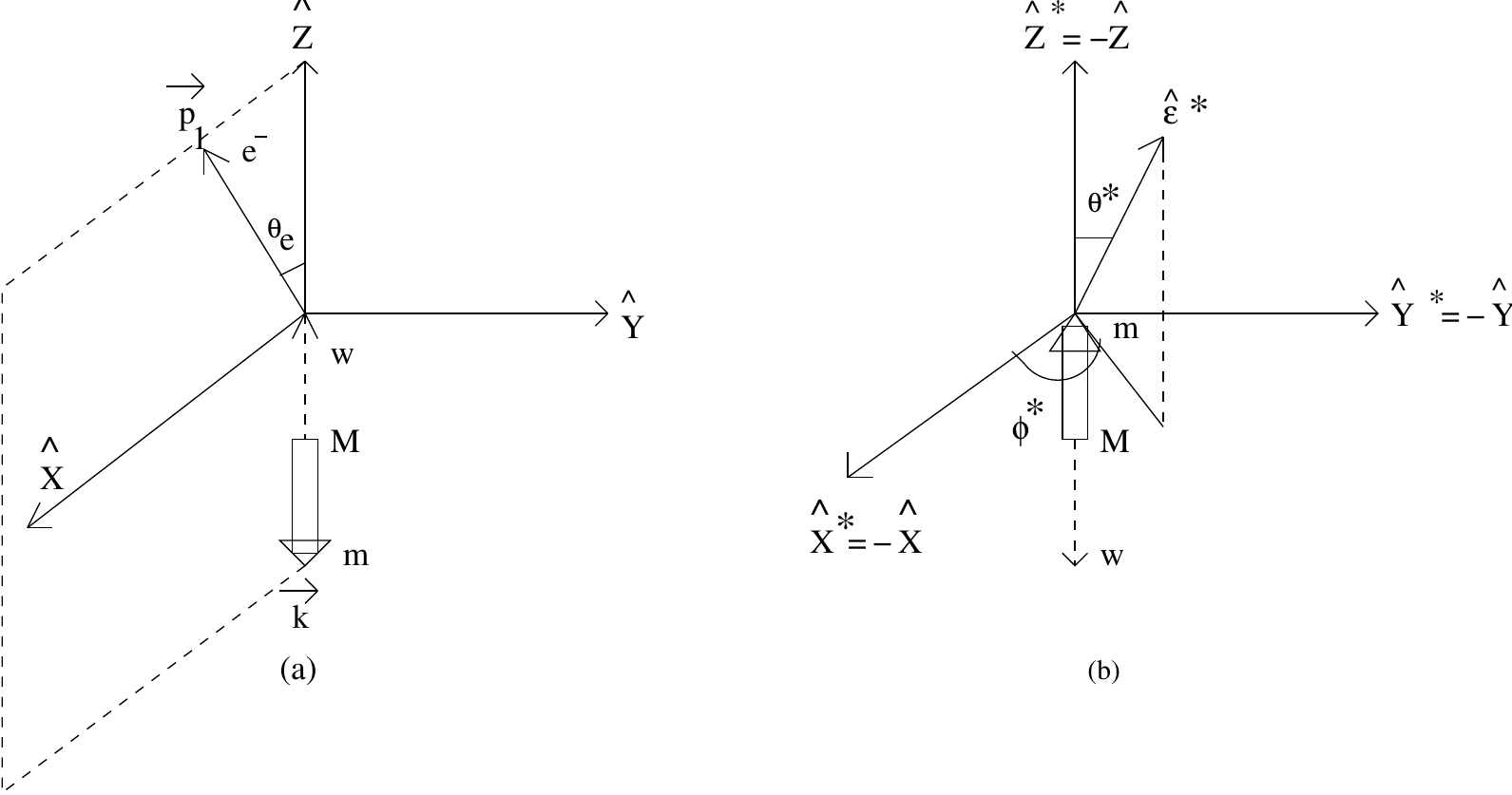}
	\end{center}
	\caption{\bf{Coordinate system for the semileptonic decay of $B_c$-meson (a) the
			decaying virtual W and (b) the decaying final vector meson}}.
\end{figure}
We introduce here a pair of dimensionless variables (y, x) scaled to the parent meson mass as $y=(q^2/{M}^2)$ and $x=(p_1.p_2/{M}^2)$ for sake of convenience to describe kinematics of the decay process. In the vanishing lepton mass limit, kinematically allowed range of y is
\begin{equation}
0\leq y\leq \left (1-{m\over M}\right )^2
\end{equation}

We also consider two frames of reference (i) parent meson rest frame and (ii)
the e$\nu_e$ center-of-mass frame. The co-ordinate system chosen here is such that the daughter meson momentum $\vec k$ is along the negative Z-axis with the charged lepton momentum
$\vec p_1$ subtending an angle $\theta _e$ to Z-axis [Fig. 1(a)] in the
e$\nu_e $ center-of-mass frame. The Y-axis is oriented perpendicular to the plane containing final particles' momenta.\\

The physical quantities of interest associated with the final state particles are their energy and momentum, which can be calculated in both frames considered here. In the e$\nu_e$
center-of-mass frame they are 
\begin{equation}
E_1  =  E_2={M\over 2}{\sqrt y}
\end{equation}
\begin{equation}
E_X  =  {M\over{2\sqrt y}}\left   (1-{{m}\over{M^2}}-y\right )
\end{equation}
\begin{equation}
\mid {\vec  k}\mid = K/{\sqrt y}
\end{equation}
\begin{equation}
K={M\over 2}\left [\left (1-{{m^2}\over{M^2}}-y\right )^2-
4{{m^2}\over{M^2}}y\right ]^{1/2}
\end{equation}
In the parent meson rest frame, however, the quantities are
\begin{equation}
{\tilde E}_1 = Mx = {K\over 2}\cos \theta _e +{M\over 4}\left 
(1-{{m^2}\over{M^2}}+y\right ).
\end{equation}
\begin{equation}
{\tilde E}_X  =  {M\over 2}\left (1+{{m^2}\over{M^2}}-y\right )
\end{equation}
\begin{equation}
\mid {\tilde {\vec k}}\mid = K
\end{equation}
The hardronic amplitudes are covariantly expanded in terms 
of lorentz invariant form factors. For the transition of type 
($0^- \to 0^-$), the expansion is 
\begin{equation}
<X(k)\mid V_{\mu}(0)\mid B_c(p)>=f_+(q^2)(p+k)_{\mu}+f_-(q^2)(p-k)_{\mu}
\end{equation}
For $(0^-\to 1^-)$ type transitions they are
\begin{equation}
<X(k,\epsilon ^*)\mid V_{\mu}(0)\mid B_c(p)> = ig(q^2)\in _{\mu \nu \rho 
	\sigma }\epsilon ^{*\nu }(p+k)^{\rho }(p-k)^{\sigma }
\end{equation}
\begin{eqnarray}
<X(k,\epsilon ^*)\mid A_{\mu}(0)\mid B_c(p)> = f(q^2)\epsilon ^*_{\mu} +
a_+(q^2)(\epsilon^*.p)(p+k)_{\mu}\nonumber\\
-a_-(q^2)(\epsilon^*.p)(p-k)_{\mu}
\end{eqnarray}  
Here $\epsilon ^*\equiv (\epsilon ^*_0,{\vec \epsilon}\;^*)$ with
$\epsilon ^*.k=0$, represents the vector meson polarization.

The differential decay rate is written in the generic form
\begin{equation}
d\Gamma=\frac{1}{2E_{B_c}}\sum_{\delta_1,\delta_2,\lambda}|{\cal M}|^2d\Pi_3
\end{equation}
where the three body phase space factor is 
\begin{equation}
d\Pi_3=(2\pi)^4\delta^{(4)}(p-p_1-p_2-k)\frac{d^3\vec k}{(2\pi)^32E_X}
\frac{d^3\vec p_1}{(2\pi)^32E_1}\frac{d^3\vec p_2}{(2\pi)^32E_2}
\end{equation}
and the invariant transition amplitude squared is given by
\begin{equation}
\sum_{\delta_1, \delta_2, \lambda}|{\cal M}|^2=\frac{G_F^2}{2}
|V_{bc}|^2L^{\mu\sigma}H_{\mu\sigma} \label{lopa}
\end{equation}
We write $L^{\mu\sigma}=\sum_{\delta_1,\delta_2}(L^\mu {L^\sigma}^\dagger)$ 
representing a sum over lepton spin indices ($\delta_1, \delta_2$)
and also  $H_{\mu\sigma}=\sum_{\lambda}(H_\mu H_\sigma^\dagger)$
representing a sum over daughter meson (vector) polarization index $\lambda$.

It is convenient to calculate the Lorentz invariant leptonic piece $L^{\mu\sigma}$ obtained in the form:
\begin{equation}
L^{\mu\sigma}=8\Biggl[(p_1^\mu p_2^\sigma - p_1.p_2 g^{\mu\sigma}+
p_1^\sigma p_2^\mu)+i\epsilon_{\mu\alpha\sigma\beta} p_1^\sigma p_2^\beta\Biggr]
\end{equation}
in the e$\nu_e$ center-of-mass frame. Since its timelike component $L^{00}$ is zero in the vanishing lepton mass limit, the non-vanishing contribution to ${\cal M}$ comes from the product $L^{ij}H_{ij}$ only. 
Then integrating $L^ij$ over the lepton phase space, one gets, in the e$\nu_e$ center-of-mass frame:
\begin{equation}
\int\int\frac{d^3{\vec p}_1}{2E_1}\frac{d^3{\vec p}_2}{2E_2}\;L^{ij}\;
\delta^{(4)}(p-p_1-p_2-k)=\frac{4\pi}{3}\;q^2\;\delta^{ij}
\end{equation}which reduces the effective hadronic part $\sum_\lambda H_{ij}$ to $\sum_\lambda H_{ii}$. With this consideration, the expression of the differential decay rate in the e$\nu_e$ 
center-of-mass frame, is transformed to the form 
\begin{equation}
d{\bar \Gamma}=\frac{1}{(2\pi)^5}\frac{1}{2E_{B_c}}\frac{G_F^2}{2}
|V_{bc}|^2\frac{d^3\vec k}{2E_X}\frac{4\pi}{3}q^2\sum_\lambda H_{ii}
\end{equation}

It is worthwhile to note here that the hadronic amplitude `$h_i$' can be expressed, in this frame, in a simple and convenient form as the terms involving the form factors $f_-(q^2)$ and $a_-(q^2)$ 
do not contribute to $\vec h$ pertaining to transitions of the type 
($0^- \to 0^-$) and ($0^- \to 1^-$), respectively. For ($0^- \to 0^-$) 
type transitions one obtains $h_i$ (12) in terms of a 
single form factor $f_+(q^2)$ as
\begin{equation}
\vec h =(\vec p+\vec k )f_+(q^2)
\end{equation}
Similarly for transitions of type $(0^-\to 1^-)$, we obtain $\vec h$ from Eq.(13) and (14) as
\begin{equation}
\vec h=2i{\sqrt y}Mg(q^2)({\vec \epsilon}\;^*\times \vec k)-f(q^2)
{\vec \epsilon}\;^*-2(\epsilon^*.p)a_+(q^2)\vec k
\end{equation}  
For the calculation of physical quantities, it is more convenient to use helicity amplitudes, which are linearly related to the invariant form factors \cite{A54,A55,A61}. We therefore,  expand $\vec h$ in
terms of helicity basis (effectively of the virtual W) as 
\begin{equation}
\vec h={\cal H}_+{\hat e}_++{\cal H}_-{\hat e}_-+{\cal H}_0{\hat e}_0
\end{equation} 
with 
\begin{equation}
{\hat e}_{\pm}={1\over{\sqrt 2}}(\mp {\hat x}-i{\hat y});\;
{\hat e}_0={\hat z}
\end{equation}
The polarization vector $\hat \epsilon^*$ with the polar and azimuthal angle
$(\theta^*,\phi^*)$ in the vector meson helicity frame [Fig. 1(b)] can be
lorentz-transformed to the (e$\nu_e$) center-of-mass frame to be obtained in the form
\begin{equation}
{\hat \epsilon}^*={1\over{\sqrt 2}}\sin\theta^*e^{i\phi^*}{\hat e}_+
-{1\over{\sqrt 2}}\sin\theta^*e^{-i\phi^*}{\hat e}_-
-{{E_X}\over{M_X}}\cos\theta^*{\hat e}_0
\end{equation}
Then expanding $h_i$ in terms of helicity basis (23, 24) and using 
the Lorentz transformed form of $\hat \epsilon^{\star}$ (25), one can obtain 
the helicity amplitudes ${\cal H}_{\pm}$ and ${\cal H}_0$ from Eq. (22) as
\begin{equation}
{\cal H}_{\pm}=\mp\frac{sin\theta^*}{\sqrt 2}e^{\pm i\phi^*}{\bar { H}}_{\pm}
\end{equation}
\begin{equation}
{\cal H}_0=cos\theta^*{\bar {H}}_0
\end{equation}
where ${\bar H}_{\pm}$ and ${\bar H}_0$ are reduced helicity amplitudes. For ($0^- \to 1^-$)-type semileptonic transitions, these reduced helicity amplitudes are obtained in terms of 
invariant form factors f, g and $a_+$ as
\begin{equation}
{\bar H}_{\pm} = [\; f(q^2)\mp 2MKg(q^2)\; ],
\end{equation}
\begin{equation}
{\bar H}_0={M\over{2m\sqrt y}}\left [\left (1-{{m^2}\over{M^2}}-y\right )f(q^2)
+4K^2a_+(q^2)\right ].
\end{equation} 
Now $H_{ii}=h_ih_i^\dagger$ in Eq. (20) can be expressed in terms of reduced helicity amplitudes (28, 29). Then integration over the polar and azimuthal angles ($\theta^*, \phi^*$) and then sum over the daughter meson (vector) polarization yield an invariant expression for the differential decay rate. Once obtained in an invariant form it is then convenient to cast in any frame (here the parent meson rest frame) so as to get the final expression of the differential decay rate as:
\begin{equation}
{{d{\tilde \Gamma}}\over{dy}}=\frac{1}{96\pi^3}{G_F}^2\mid V_{bc}\mid^2M^2Ky
\;[\;\mid {\bar H}_+\mid^2+\mid {\bar H}_-\mid^2+\mid {\bar H}_0\mid^2\;]
\end{equation}
Here the contribution of $\mid {\bar H}_0\mid^2$ term to the 
differential decay rate (30) refers to the longitudinal mode and that of the combined term $[\;\mid {\bar H}_+\mid^2+\mid {\bar H}_-\mid^2\;]$
refers to the transverse polarization mode for the semileptonic transitions of the type ($0^- \to 1^-$). However, in case of ($0^- \to 0^-$) type transitions, one can realize corresponding expressions by appropriately identifying
\begin{equation}
\bar H_{\pm}=0;\;\;\bar H_0=-2{K\over{\sqrt y}}f_+(q^2)
\end{equation}
which leads to the differential decay rate in parent meson  
rest frame as
\begin{equation}
{{d{\tilde \Gamma}}\over{dy}}={{{G_F}^2\mid V_{bc}\mid^2K^3M^2}\over{24\pi^3}}
\mid f_+(q^2)\mid^2
\end{equation}

\section
{Transition Matrix Element and Weak Form Factors}
The decay process physically occurs between the momentum eigenstates of participating mesons. Therefore in a field theoretic description they need to be represented by their appropriate momentum wave packets reflecting the momentum and spin distribution between constituent quark and antiquark inside the respective meson core. In the RIQ model, the appropriate wave packet representing the meson state $|B_c(\vec p,S_{B_c})>$ is consideed at definite momentum $\vec p$ and spin state $S_{B_c}$ as \cite{A52,A53,A54,A55,A56,A57,A58,A59} 
\begin{equation}
|B_c(\vec p, S_{B_c})>={\hat \Lambda(\vec p,S_{B_c})}|(\vec p_{b},\lambda_b);(\vec p_{c},\lambda_c)>
\end{equation}
where, $|(\vec p_{b},\lambda_b);(\vec p_{c},\lambda_c)>= \hat b_{b}^\dagger (\vec p_{b},\lambda_b) 
{\hat {\tilde{b}}}_{c}^\dagger (\vec p_{c},\lambda_{c})|0>$ is a fockspace representation of the unbound quark and antiquark in a 
color-singlet configuration with their respective momentum and spin as  $(\vec p_{b},\lambda_{b})$ and $(\vec p_{c},\lambda_{c})$. 
Here $\hat b_{b}^\dagger (\vec p_{b},\lambda_{b})$ and  ${\hat {\tilde{b}}}_{c}^\dagger (\vec p_{c},\lambda_{c})$
are, respectively the quark and antiquark creation operators. ${\hat \Lambda}(\vec p,S_{B_c})$ represents an integral operator: 
\begin{equation}
{\hat \Lambda}(\vec p,S_{B_c})=\frac{\sqrt 3}{\sqrt{N_{B_c}(\vec p)}}\;
\sum _{{\delta_b},{\delta_{\bar c}}}\zeta_{bc}^{B_c}(\lambda_{b},\lambda_{c})
\int d^3{\vec p}_{b}\;d^3{\vec p}_{c}\;\delta^{(3)}(\vec p_{b}+\vec p_{c}-\vec p)
{\cal G}_{B_c}(\vec p_{b},\vec p_{c})
\end{equation}
Here $\sqrt 3$ is the effective color factor, 
$\zeta^{B_c}(\lambda_{b},\lambda_{c})$
stands for SU-(6) spin flavor coefficients for the $B_c$ meson.
$N(\vec p)$ is the meson-state normalization which is realized from
$<{B_c}(\vec p)\mid {B_c}({\vec p}\;^{\prime})>=\delta ^{(3)}(\vec p-{\vec p}\;^{\prime})$ 
in an integral form
\begin{equation}
N(\vec p)=\int d^3{\vec p}_{b}\;\mid {\cal G}_{B_c}({\vec p}_{b},\vec p-{\vec p}_{b})\mid ^{2}
\end{equation}

Finally, ${\cal G}_{B_c}({\vec p_{b}}, \vec p - \vec p_{b})$ is the effective momentum profile function for the quark-antiquark pair. In terms of individual momentum probability amplitudes ${G}_{b}(\vec p_{b})$ and ${G}_{c}(\vec p_{c})$ of the constituent quark b and c, respectively, ${\cal G}_{B_c}({\vec p_{b}}, \vec p - \vec p_{b})$ is taken in this model in the  form:
\begin{equation}
{\cal G}_{B_c}({\vec p_{b}},{\vec p_{c}})=\sqrt{G_{b}(\vec p_{b}){\tilde G}_{c}(\vec p_{c})}
\end{equation}
in a straight forward extension of the ansatz of Margolis and Mendel in their bag model analysis \cite{A62}. A brief account of the model framework and quark orbitals derivable in the RIQ model along with those of the corresponding momentum probability amplitudes are given in the Appendix.
In the wave packet representation of meson bound state $|B_c(\vec p,S_{B_c})>$, the bound state character is infact embedded in the effective momentum distribution function ${\cal G}_{B_c}({\vec p_{b}}, \vec p_{c})$ . Any residual internal dynamics responsible for decay process can therefore be described at the level of otherwise free quark and antiquark using appropriate Feynman diagram shown in Fig. 2. The total contribution from the Feynman diagram provides the constituent level S-matrix element $S_{fi}^{bc}$  which when operated by the baglike operator gives the meson level effective S-matrix element $S_{fi}^{B_c}$  as:
\begin{equation}
S_{fi}^{B_c} = {\hat \Lambda} S_{fi}^{bc}
\end{equation}
\begin{figure}
	\begin{center}
		\includegraphics[width=13cm,height=8cm]{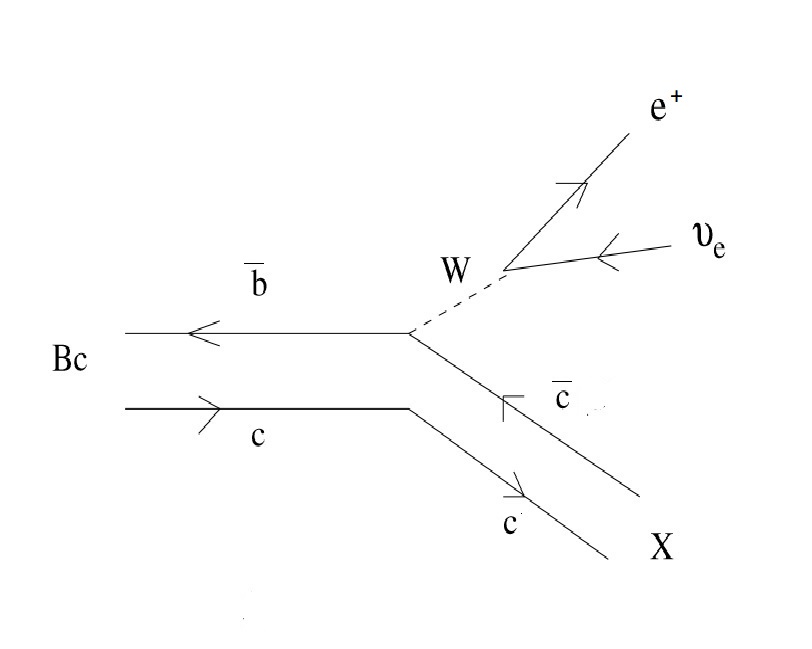}
	\end{center}
	\caption{\bf{Semileptonic decay of $B_c$ meson}}.
	\end{figure}

\subsection{Transition amplitude}
The S-matrix element for the decay for $B_c^{+} \to X e^+\nu_e$  depicted in Fig.2 can be written in general form: 
\begin{equation}
S_{fi}=-i\frac{G_F V_{bc}}{\sqrt 2 (2\pi)^4}\int d^4{x_1}d^4{x_2}d^4 qe^{-iq(x_2 - x_1)} <e^+(\vec p_{1}, \delta_{1})\nu(\vec p_{2}, \delta_{2})|J^{\mu}_l(x_2)J^{\mu}_l(x_1)|B_c(\vec p, S_{B_c})>
\end{equation}
Here we consider the parent meson $B_c$ in the ground state and the daughter meson in either $X(nS)$ state. The matrix element corresponding to leptonic weak current is found as
\begin{equation}
<e^+(\vec p_{1}, \delta_{1})\nu_e(\vec p_{2}, \delta_{2})|j^{\mu}_l(x_2)|0>= \frac{e^{i(\vec p_1 + \vec p_2)x_2}}{\sqrt {(2\pi)^3 2E_1(2\pi)^32E_{2}}} l^{\mu}
\end{equation}

with $l^{\mu}={\bar u}_e(\vec p_1, \delta_1)\gamma^{\mu}(1-\gamma_5)v_{\nu_e}
(\vec p_2, \delta_2)$

Using the appropriate meson states the relevant hadronic matrix element can be obtained in the form:
\begin{eqnarray}
&<X(\vec k,S_X)|{J_\mu}{^h}(x_1)|B_c(\vec p, S_{B_c})>=
\frac{(2\pi)^4}{\sqrt{N_{B_c}(\vec p)N_X(\vec k)}}
\int\frac{d^3\vec p_{b}}{\sqrt{(2\pi)^32E_{\vec p_{b}}(2\pi)^32E_{\vec k+\vec p_{b}-\vec p}}}\nonumber\\
&\times{\cal G}_{B_c}(\vec p_{b},\vec p-\vec p_{b}){\cal G}_X(\vec k+\vec p_{b}-\vec p, \vec p-\vec p_{b})\times <S_X|J_\mu|S_{B_c}>
\end{eqnarray}
Here $E_{\vec p_{b}}$ and $E_{\vec p_{b}+\vec k-\vec p}$ 
stand for the energy of non-spectator quark of the parent and daughter meson, respectively and $<X(\vec k,S_X)|{J_\mu}{^h}(0)|B_c(\vec p, S_{B_c})>$ represents symbolically the spin matrix elements of vector-axial vector current.
Using matrix element of leptonic and hadronic weak current we obtain S-matrix element for the decay in the standard form:
\begin{equation}
S_{fi}=(2\pi)^4\delta^{(4)}(p-k-p_1-p_2)(-i{\cal M}_{fi})\times \frac{1}{\sqrt{{(2\pi)^3} 2E_{B_c}}}\prod_f\Biggl(\frac{1}{\sqrt{2E_f {(2\pi)^3}}}\Biggr)
\end{equation} where the the invariant transition amplitude ${\cal M}_{fi}$ is:
\begin{center}
${\cal M}_{fi}=\frac{G_F}{\sqrt 2} V_{bc}l^\mu h_{\mu}$
\end{center}
Here the hadronic amplitude $h_\mu$ is obtained in the 
parent $(B_c)$-meson rest frame in the form
\begin{eqnarray}
{h_\mu}=\frac{\sqrt{4ME_X}}{\sqrt{N_{B_c}(0)N_X(\vec k)}}
\int \frac{d^3\vec p_{b}}{\sqrt{2E_{\vec p_{b}}2E_{\vec k+\vec p_{b}}}}
&\times&{\cal G}_{B_c}(\vec p_{b},-\vec p_{b}){\cal G}_X
(\vec k+\vec p_{b}, -\vec p_{b})\nonumber\\
&\times&<S_X|J_\mu(0)|S_{B_c}>
\end{eqnarray}

\subsection{Weak decay form factors}
In order to extract model expression of weak decay form factors, the covariant exapansion of hadronic decay amplitudes \cite{A12,A13,A14} are compared with corresponding model expression . For this the spin matrix element are simplified as follows. 
For $0^-\to 0^-$ transitions, the axial vector current does not contribute. The non-vanishing vector current parts are obtained as:

\begin{equation}
<S_X(\vec k)|V_0|S_{B_c}(0)>=\frac{(E_{\vec p_b+\vec k}+
	m_c)(E_{\vec p_b}+m_b)+\vec p_b\;^2}{\sqrt{(E_{\vec p_b+\vec k}+m_c)
		(E_{\vec p_b}+m_b)}}
\end{equation}
\begin{equation}
<S_X(\vec k)|V_i|S_{B_c}(0)>=\frac{(E_{\vec p_b}+m_b)k_i}
{\sqrt{(E_{\vec p_b+\vec k}+m_c)(E_{\vec p_b}+m_b)}}
\end{equation}
\newpage 

With these results the form factor $f_+(q^2)$ for $0^- \to 0^-$ type transitions is found in the form:
\begin{equation}
f_+=\frac{1}{2M}\int d\vec p_{b} {\cal C}(\vec p_{b})[(E_{p_b}+m_b)
(E_{\vec p_{b}+\vec k}+M-{E_X})+\vec p_{b}\;^2] 
\end{equation}
where
\begin{eqnarray}
{\cal C}(\vec p_{b}) = \sqrt{\frac{M{E_X}}
	{N_{B_c}(0)N_X(\vec k)}}
\times\frac{{\cal G}_{B_c}(\vec p_{b}, -\vec p_{b})
	{\cal G}_X(\vec k+\vec p_{b}, -\vec p_{b})}{\sqrt{E_{\vec p_{b}}E_{\vec p_{b}+\vec k}(E_{\vec p_{b}}+m_b)(E_{\vec p_{b}+\vec k}+m_c)}}
\end{eqnarray}
However for $(0^- \to 1^-)$ transitions, the spin matrix element for vector and axial vector current are obtained separately as: 
\begin{equation}
<S_X(\vec k, {\hat\epsilon^*})|V_0|S_{B_c}(0)>=0
\end{equation}
\begin{equation}
<S_X(\vec k,{\hat\epsilon^*})|V_i|S_{B_c}(0)>=\frac
{i(E_{\vec p_{b}}+m_b)({\hat\epsilon^*}\times\vec k)_i}
{\sqrt{(E_{\vec p_{b}}+m_b)(E_{\vec p_{b}+\vec k}+m_c)}}
\end{equation}
\begin{equation}
<S_X(\vec k, {\hat\epsilon^*})|A_i|S_{B_c}(0)>=\frac{[(E_{\vec p_{b}}+m_b)(E_{\vec p_{b}+\vec k}+m_c)-\vec p_{b}\;^2/3]\epsilon^*_i}
{\sqrt{(E_{\vec p_{b}}+m_b)(E_{{p_b}+\vec k}+m_c)}}
\end{equation}
\begin{equation}
<S_X(\vec k,{\hat\epsilon^*})|A_0|S_{B_c}(0)>=
\frac{-(E_{\vec p_{b}+m_b})({\hat\epsilon^*}.\vec k)}{\sqrt{(E_{\vec p_{b}}+m_b)(E_{\vec p_{b}+\vec k}+m_c)}}
\end{equation}
\\
A term by term comparison of results in Eqs.(48-49) with corresponding expressions from Eqs.(13-14) yields the form-factor $g(q^2)$ and $f(q^2)$ in the form:

\begin{equation}
g=-\frac{1}{2M}\int d\vec p_{b}{\cal C}
(\vec p_{b})(E_{\vec p_{b}}+m_b)
\end{equation}\
\begin{equation}
f=-\int d\vec p_{b}{\cal C}^0(\vec p_{b})
[(E_{\vec p_{b}}+m_b)(E_{\vec p_{b}+\vec k}+m_c) - \vec p_{b}^2/3]
\end{equation} where
\begin{eqnarray}
{\cal C}^0(\vec p_{b}) = \sqrt{\frac{Mm}
	{N_{B_c}(0)N_X(0)}}
\times\frac{{\cal G}_{B_c}(\vec p_{b}, -\vec p_{b})
	{\cal G}_X(\vec p_{b}, -\vec p_{b})}{\sqrt{E_{\vec p_{b}}E^{0}_{\vec p_{b}} (E_{\vec p_{b}}+m_b)(E_{\vec p_{b}}+m_c)}}
\end{eqnarray}
with, $E^0_{p_b}=\sqrt{{|\vec p_b|^2}+m_c^2}$. 

Now considering both the timelike and spacelike parts of axial vector current contribution and simplifying, we finally obtain the model expression for weak form factor $a_+(q^2)$ in the form:
\begin{equation}
a_+=-\frac{1}{2M^2}\Big[(J - f)+\frac{(I-{f})E_X(M-{E_X})}{{E_X}^2-{m^2}}\Big]
\end{equation}
where,
\begin{eqnarray}
J = \sqrt{\frac{M{E_X}}
	{N_{B_c}(0)N_X(\vec k)}}
\times\int \frac{d\vec p_b{\cal G}_{B_c}(\vec p_{b}, -\vec p_{b})
	{\cal G}_X(\vec p_{b}+\vec k, -\vec p_{b})(E_{\vec p_{b}} + m_{b})E_X}{\sqrt{E_{p_b}E_{\vec p_{b}+\vec k} (E_{\vec p_{b}}+m_b)(E_{\vec p_{b}+\vec k}+m_c)}}
\end{eqnarray}
and
\begin{equation}
I = \sqrt{\frac{M{E_X}}
	{N_{B_c}(0)N_X(\vec k)}}
\times \int \frac{d\vec p_b{\cal G}_{B_c}(\vec p_{b}, -\vec p_{b})
	{\cal G}_X(\vec p_{b}+\vec k, -\vec p_{b})}{\sqrt{E_{p_b} E_{\vec p_{b}+\vec k} (E_{\vec p_{b}}+m_b)(E_{\vec p_{b}+\vec k}+m_c)}}
	 [(E_{\vec p_{b}}+m_b)(E_{\vec p_{b}+\vec k}+m_c) - {\vec p}_{b}\;^2/3]
\end{equation}
Note that we calculate the timelike part of axial vector current contribution corresponding to longitudinal spin polarization of daughter meson. Because the spin quantization axes is taken here opposite to the boost direction, the longitudinal polarization vector ${\epsilon_0^{*(L)}}$ is therefore boosted yielding its timelike components ${\epsilon_0^{*(L)}} = {-\vec k\over m}$ and 
${\epsilon_0^{*(T)}} = 0$
The model expressions for the form factors $f_+(q^2)$, $g(q^2)$, $f(q^2)$ and $a_+(q^2)$ in Eqs. (45), (51), (52) and (54) are believed to embody their $q^2$ dependence in the allowed kinematic range. The weak form factors can also be expressed in the dimensionless forms as cited in the literature to treat all in the same footing as
\begin{eqnarray}
F_1(q^2)&=&f_+(q^2)\nonumber\\
V(q^2)&=&(M_{B_c}+M_X)g(q^2)\nonumber\\
A_1(q^2)&=&(M_{B_c}+M_X)^{-1}f(q^2)\nonumber\\
A_2(q^2)&=&-(M_{B_c}+M_X)a_+(q^2)
\end{eqnarray}
The $q^2$- dependence of the weak form factors and prediction of the branching ratios for semileptonic $B_c$-decays to S-wave charmonium states can then be obtained using relevant hadronic quantities and RIQ-model parameters, as described in the next section.

\newpage 

\section{Numerical Results and Discussion}
For numerical calculation, we take the model parameters (a,$V_0$), quark masses $m_q$ and quark binding energies $E_q$ which have already been fixed from hadron spectroscopy by fitting the data of heavy flavored mesons, and used earlier to describe wide ranging hadronic phenomena: \cite {A52,A53,A54,A55,A56,A57,A58,A59}
\begin{eqnarray}
(a, V_0)&\equiv & (0.017166\;{GeV}^3, -0.1375\;GeV)\nonumber\\
(m_{b}, m_{c},E_{b}, E_{c})&\equiv & (4.77659, 1.49276, 4.76633, 1.57951)\;GeV
\end{eqnarray}
To obtain the binding energy of constitute quarks (b,c) in radially excited (2S and 3S) meson states, we solve the cubic equations that represent the corresponding independent quark bound state condition. Accordingly we take \cite{A58,A59}
\begin{eqnarray}
( E_{b} ; E_{c} ) = ( 5.05366 ; 1.97016 ) GeV \nonumber\\
( E_{b} ; E_{c} ) = ( 5.21703 ; 2.22479 ) GeV 
\end{eqnarray}
The masses of participating mesons in their S-wave ground and radially excited states are taken in GeV as\cite{A59}
$M_{B_c}$=6.2749, $M_{\eta_c}$=2.9839, $M_{\eta_c(2S)}$=3.6392, 
$M_{\eta_c(3S)}$=3.8381, $M_{J/\psi}$=3.0968, $M_{\psi(2S)}$=3.6860, $M_{\psi(3S)}$=4.1104. For CKM parameter $V_{bc}$ and $B_c$ meson lifetime $\tau_{B_c}$, we use their central values from Particle Data Group \cite{A60} as $V_{bc}$=0.0422 and $\tau_{B_c}=0.51^{+0.18}_{-0.16}(stat.)\pm 0.03(syst.)\ ps$. 
Before calculating the weak form factors and their $q^2$- dependence in the allowed kinematic range with the input parameters \cite{A58,A59}, it is interesting to study the behavior of radial quark momentum distribution amplitude function related to $B_c$ meson state together with those of the final S-wave charmonium states. The shape of the behavior of momentum distribution amplitude is shown in Fig.3. One can see that the overlap region between the momentum distribution amplitude function for the initial $B_c$ meson state and final charmonium 1S state is maximum, where as it is less for $B_c$ decay mode to 2S and least for the $B_c$ decay to 3S charmonium state. The lorentz invariant weak form factors representing the decay amplitudes are in fact calculated from the overlapping integrals of participating hadron wave functions. It is evident therefore that the contribution of weak form factors to the decay width/branching fractions should be obtained in the decreasing order of magnitude as one considers various semileptonic $B_c$ decays to S-wave charmonium states from 1S to higher 2S and 3S states. 
\begin{figure}[!htb]
	\begin{subfigure}{1.30\textwidth}
		\includegraphics[width=\linewidth]{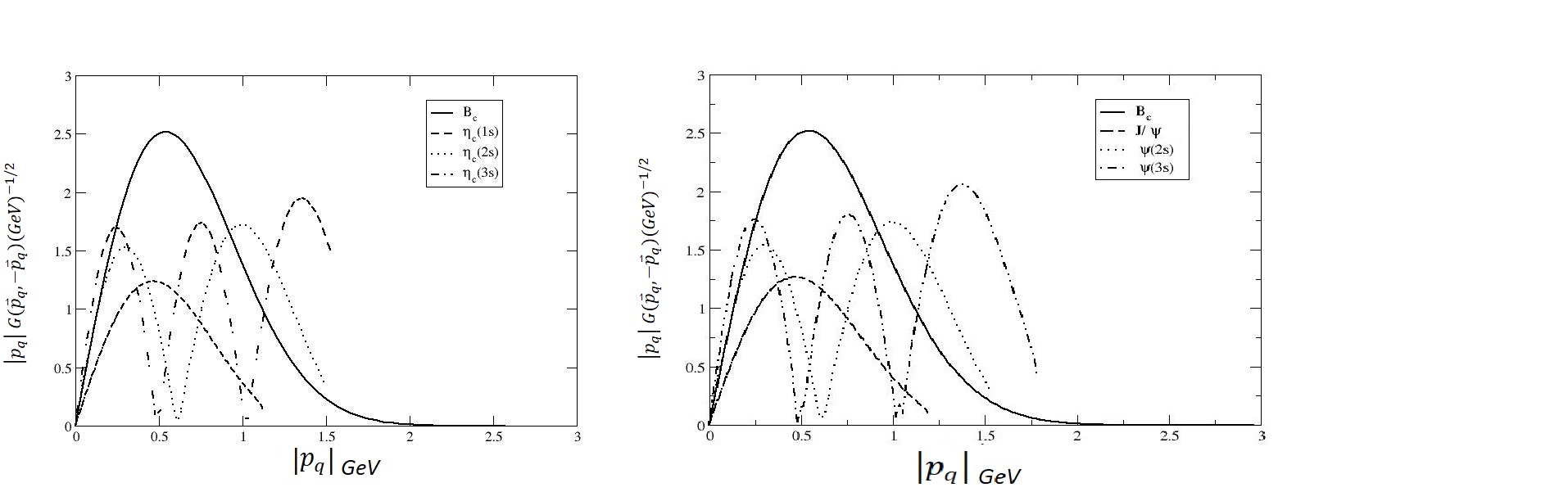}
		
	\end{subfigure}

	\caption{The overlap of momentum distribution amplitudes of initial and final meson state.}
\end{figure}
Our predicted $q^2$- dependence of weak form factors for six decay modes: in their physical kinematic range is shown in Fig 4. We find that transitions $B_c^+ \to \eta_c,(J/\psi) e^+ \nu_e$ have a relatively strong $q^2$-dependence as relevant form factors become larger with increasing $q^2$. This behavior, however is not universal. For example,  for transition, $B_c^+ \to \eta_c(2S), \psi(2S) e^+\nu_e$ and $B_c^+ \to \eta_c(3S), \psi(3S) e^+\nu_e$ some of the form factors decreases with increasing $q^2$. Similar predictions have been made in other model calculations based on perturbative QCD approach \cite{A40} light-front quark model \cite{A36} and ISGW2  quark model \cite{A34}. This is attributed to the nodal structure in the momentum distribution amplitude functions corresponding to $B_c$ decay to different S- wave charmonium states and the momentum transfer involved in different decay modes.
\begin{figure}[!htb]
	\begin{subfigure}{0.48\textwidth}
		\includegraphics[width=\linewidth]{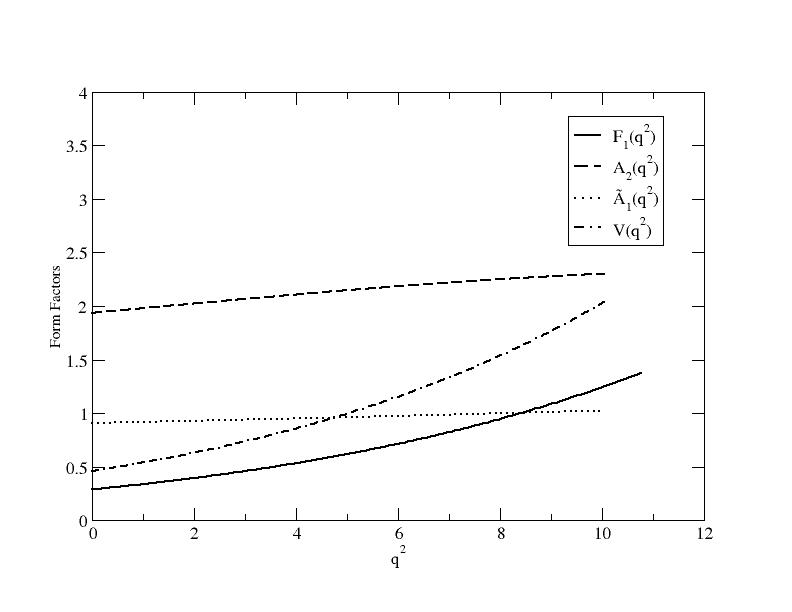}
		\caption{$B_c\to X(1S) e \nu_e$} 
	\end{subfigure}
	\begin{subfigure}{0.48\textwidth}
		\includegraphics[width=\linewidth]{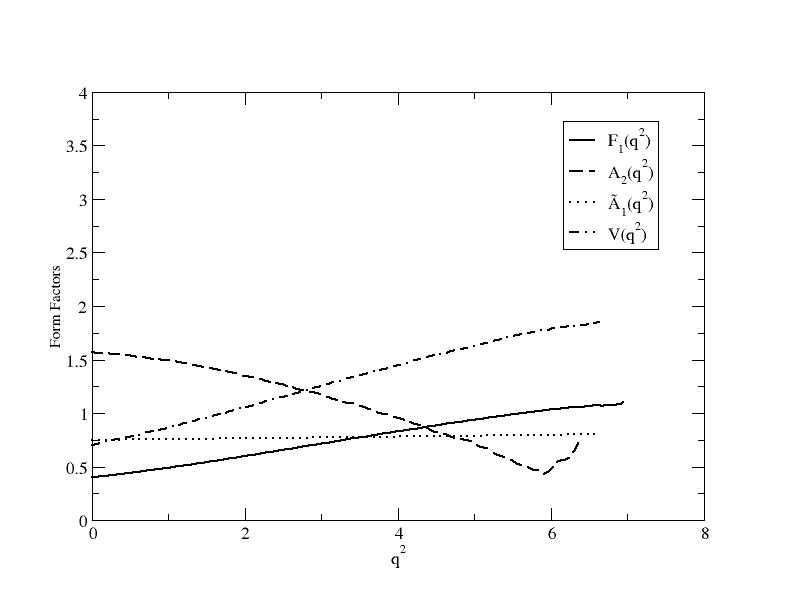}
		\caption{$B_c\to X(2S) e \nu_e$} 
	\end{subfigure}
	\begin{subfigure}{0.48\textwidth}
		\includegraphics[width=\linewidth]{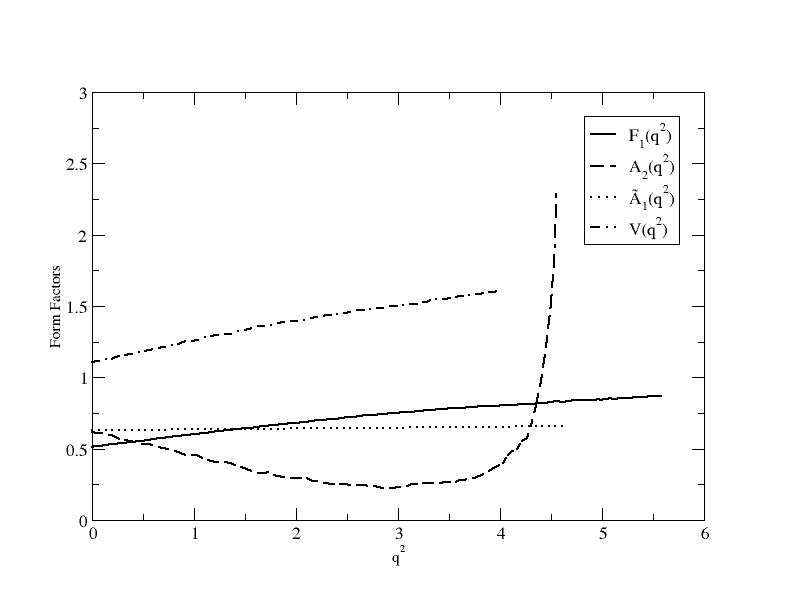}
		\caption{$B_c\to X(3S) e\nu_e$} 
	\end{subfigure}
	\caption{The $q^2$- dependence of form factors of $B_c\to X e \nu_e$} 
\end{figure}
\newpage 
One may naively expect the weak form factors to satisfy the heavy quark symmetry (HQS) relations: 
\begin{equation}
F_1(q^2)\simeq V(q^2)\simeq A_2(q^2)\simeq \tilde{A_1}(q^2)
\end{equation}
with \begin{center}{$\tilde{A_1}(q^2)=
[1-\frac{q^2}{(M+m)^2}]^{-1} A_1(q^2)$}.
 \end{center}
as an outcome of heavy quark effective theory (HQET). From the predicted $q^2$- dependence (Fig.4), it is evident that the weak form factors do not simultaneously satisfy HQS relation. This corroborates to the well known fact that the HQS is not strictly applicable to the case of mesons with two heavy quarks. Integrating the expression for partial decay width over the allowed kinematic range of y = $\frac{q^2}{M^2}$, one can calculate the decay width and hence BR for six decay modes considered in this work. The results of evaluation of BR for all considered decays are shown in Tabel-I in comparison with other model predictions.\\

\begin{table}[ht]
	\renewcommand{\arraystretch}{1}
	\setlength\tabcolsep{2.0pt}
	\begin{center}
		\caption{{\bf Predicted branching ratios $(\%)$ of $B_c^+\to X e^+\nu_e$ 
				decays in comparison with other model predictions }}
		\begin{tabular}{|c|c|c|c|c|c|c|c|c|c|c|c|}
			\hline
			\hline
			Transition & Our work & \cite{A20} & \cite{A23,A24} & \cite{A25} & \cite{A27} & \cite{A29,A30} & \cite{A32} & \cite{A33}& \cite{A34} &\cite{A37}& \cite{A40}  \\
			\hline
			$B_c^+\to \eta_c e^+\nu_e$ & 0.397 & 2.1 & 0.55 & 0.42 & 0.81 & ... & 1.64 & 0.67 & 0.48 & 0.5&4.5\\
			$B_c^+\to J/\psi e^+\nu_e$ & 2.299 & 6.7 & 1.73 & 1.23 & 2.07 & ... & 2.37 & 1.49 & 1.54 & 3.3&5.7\\
			$B_c^+\to \eta_c(2S)e^+\nu_e$ & 0.236 & ... & 0.07 & 0.03 & ... & 0.11 & ... & ... & ... & 0.02&0.77\\
			$B_c^+\to \psi(2S)e^+\nu_e$ & 0.862 & ... & 0.1 & 0.03 & ... & ... & ... & ... & ... & 0.12&1,2\\
			$B_c^+\to \eta_c(3S)e^+\nu_e$ & 0.189 &... &... & $5.5\times 10^{-4}$ & ...  & $1.9\times 10^{-2}$ & ... & ... & ... & ... & 0.14\\
			$B_c^+\to \psi(3S)e^+\nu_e$ & 0.353 &...  & ... & $5.7\times 10^{-4}$ & ... & ... & ... & ... & ... & ...&$3.6\times 10^-{2}$ \\
			\hline
			\hline
		\end{tabular}
	\end{center}
\end{table}
For $B_c \to \eta_c/(J/\psi) e \nu_e$ decays, our predictions are in good agreement with those of non-relativistic quark model \cite{A20}, relativistic quark model \cite{A23,A24,A25}, Bethe-Salpeter quark model \cite{A27}, light-cone QCD sum rule \cite{A33} and QCD potential model \cite{A37}. For $B_c \to \eta_c(2S)/\psi (2S) e \nu_e$ decays our results are in overall agreement with the results of perturbative QCD approach \cite{A40}. $B_c \to \eta_c(3S)/\psi (3s) e\nu_e$ transitions have been analyzed by a few theoretical approaches \cite{A23,A24,A30,A40}. 
The order of magnitude of predicted BR in these decay modes vary widely from one model to other. While the relativistic quark model predictions \cite{A23,A24}  are typically smaller,those of the light-front QCD  sum rule \cite{A33} and perturbative QCD  approach \cite{A40} are 2 order of magnitude higher. Our predictions in this sector, although found over estimated, compared to that of \cite{A25} are in overall agreement with \cite{A40}. Since data in this sector are scant, the results of all these approaches can only be discriminated in future LHC experiments. As expected, our predicted BR are obtained in the following hierarchy:  BR($B_c^+\to \eta_c\psi (3S)$) $<$ BR($B_c^+\to \eta_c \psi(2S)$) $<$ BR($B_c^+\to \eta_c, J/\psi(1S)$).
This is due to the tighter phase space and weaker $q^2$- dependence of weak form factors contributing to decays to higher excited charmonium states. It is important to study the longitudinal ($\Gamma_L$) and transverse ($\Gamma_T$) polarization contribution to the BR of $B_c \to \psi(nS)e\nu_e$ decays in the lower, higher and whole physical region. Our predicted polarization ratios and BR in the region-I ($0\leq\ q^2\leq\frac {(M-m)^2}{2}$), region-II ($\frac {(M-m)^2}{2}\leq\ q^2\leq\ (M-m)^2$) and in whole physical region are given separately in Table II.
.
\begin{table}[ht]
	\renewcommand{\arraystretch}{1}
	\setlength\tabcolsep{3.5pt}
	\begin{center}
		\caption{{\bf The partial branching ratios $(\%)$ and polarization ratio: $\frac{\Gamma_L}{\Gamma_T}$ of $B_c^+\to Xe^+ \nu_e$ 
				decays in different $q^2$ regions }}
		\begin{tabular}{|c|c|c|c|}
			\hline
			\hline
			Transition & Region-I & Region-II & Total Region \\
			\hline
			$B_c^+\to J/\psi e^+\nu_e$ & 0.937 & 1.358 & 2.299 \\
			$\frac{\Gamma_L}{\Gamma_T}$ & 0.596 & 0.444 & 0.503 \\
			\hline 
			$B_c^+\to \psi(2S)e^+\nu_e$ & 0.390 & 0.470 & 0.862 \\
			$\frac{\Gamma_L}{\Gamma_T}$& 1.164 &0.649 &0.848 \\
			\hline 
			$B_c^+\to \psi(3S)e^+\nu_e$ & 0.190 &0.162  & 0.353 \\
			$\frac{\Gamma_L}{\Gamma_T}$ & 2.108& 0.732 & 1.276 \\
			\hline
			\hline 
		\end{tabular}
	\end{center}
\end{table}

\noindent We find the polarization ratios $\frac{\Gamma_L}{\Gamma_T} < 1$ in region-I, region-II and whole physical region which means that semileptonic $B_c$ decays to S-wave charmonium states take place predominantly in transverse mode throughout the region. This is because in our model calculation, the form factor V$(q^2)$ increases throughout with increase in $q^2$ which enhances the transverse polarization contribution in large $q^2$ region. On the other hand, form factor $A_2{(q^2)}$ which provides dominant contribution to $\Gamma_L$ as compared to $A_1{(q^2)}$ is found suppressed mostly in large $q^2$ region giving minimal contribution to $\Gamma_L$.

\newpage 
 
 \section{Summary and Conclusion}

In this paper we analyze the semileptonic $B_c$ decays to S-wave charmonium states in the framework of relativistic independent quark model based on confining potential in equally mixed scalar-vector harmonic form. The weak form factors as overlap integrals of the participating mesons' wave functions, derived from the RIQ model dynamics, are calculated explicitly in the entire kinematic range. We predict the branching ratios (BR), longitudinal to transverse polarization ratio$  $s $\frac {\Gamma_L}{\Gamma_T}$ for these decays in general agreement with predictions of other theoretical approaches. It is found the predicted BR's for $B_c$ decays to the ground state charmonium is  comparatively large $\sim$ $10^{-2}$ while those for decays to higher excited charmonium states are relatively small owing to the phase space suppression and weaker $q^2$ dependence of the form factors. The partial BR and transverse and longitudinal polarizations are investigated separately for $B_c\to \psi(nS)e\nu_e$ decays from which we find  that the ratios $\frac{\Gamma_L}{\Gamma_T}$ $<$ 1 in the lower and higher $q^2$ region as well as in the whole physical region. This means the semileptonic $B_c$ decays to S-wave charmonium vector states take place in predominantly transverse mode. These theoretical predictions could be tested in the ongoing and forthcoming experiments. With the possible data on $B_c$ decays expected from the LHC experiments one can extract the accurate value of CKM parameter which would provide an important consistency check for the standard model.  

\appendix
\setcounter{section}{0} 
\section{ CONSTITUENT QUARK ORBITALS AND MOMENTUM PROBABILITY AMPLITUDES}

In RIQ model a meson is picturised as a color-singlet assembly of a quark and an antiquark independently confined by an effective and average flavor independent potential in the form:
$U(r)=\frac{1}{2}(1+\gamma^0)(ar^2+V_0)$ where ($a$, $V_0$) are the potential parameters. It is believed that the zeroth order quark dynamics  generated by the phenomological confining potential $U(r)$ taken in equally mixed scalar-vector harmonic form can provide adequate tree level description of the decay process being analyzed in this work. With the interaction potential $U(r)$ put into the zeroth order quark lagrangian density, the ensuing Dirac equation admits static solution of positive and negative energy as: 
\begin{eqnarray}
\psi^{(+)}_{\xi}(\vec r)\;&=&\;\left(
\begin{array}{c}
\frac{ig_{\xi}(r)}{r} \\
\frac{{\vec \sigma}.{\hat r}f_{\xi}(r)}{r}
\end{array}\;\right)U_{\xi}(\hat r)
\nonumber\\
\psi^{(-)}_{\xi}(\vec r)\;&=&\;\left(
\begin{array}{c}
\frac{i({\vec \sigma}.{\hat r})f_{\xi}(r)}{r}\\
\frac{g_{\xi}(r)}{r}
\end{array}\;\right){\tilde U}_{\xi}(\hat r)
\end{eqnarray}
where, $\xi=(nlj)$ represents a set of Dirac quantum numbers specifying 
the eigen-modes;
$U_{\xi}(\hat r)$ and ${\tilde U}_{\xi}(\hat r)$
are the spin angular parts given by,
\begin{eqnarray}
U_{ljm}(\hat r) &=&\sum_{m_l,m_s}<lm_l\;{1\over{2}}m_s|
jm>Y_l^{m_l}(\hat r)\chi^{m_s}_{\frac{1}{2}}\nonumber\\
{\tilde U}_{ljm}(\hat r)&=&(-1)^{j+m-l}U_{lj-m}(\hat r)
\end{eqnarray}
With the quark binding energy $E_q$ and quark mass $m_q$
written in the form $E_q^{\prime}=(E_q-V_0/2)$,
$m_q^{\prime}=(m_q+V_0/2)$ and $\omega_q=E_q^{\prime}+m_q^{\prime}$, one 
can obtain solutions to the resulting radial equation for 
$g_{\xi}(r)$ and $f_{\xi}(r)$in the form:
\begin{eqnarray}
g_{nl}&=& N_{nl} (\frac{r}{r_{nl}})^{l+l}\exp (-r^2/2r^2_{nl})
L_{n-1}^{l+1/2}(r^2/r^2_{nl})\nonumber\\
f_{nl}&=& N_{nl} (\frac{r}{r_{nl}})^{l}\exp (-r^2/2r^2_{nl})\nonumber\\
&\times &\left[(n+l-\frac{1}{2})L_{n-1}^{l-1/2}(r^2/r^2_{nl})
+nL_n^{l-1/2}(r^2/r^2_{nl})\right ]
\end{eqnarray}
where, $r_{nl}= a\omega_{q}^{-1/4}$ is a state independent length parameter, $N_{nl}$
is an overall normalization constant given by
\begin{equation}
N^2_{nl}=\frac{4\Gamma(n)}{\Gamma(n+l+1/2)}\frac{(\omega_{nl}/r_{nl})}
{(3E_q^{\prime}+m_q^{\prime})}
\end{equation}
and
$L_{n-1}^{l+1/2}(r^2/r_{nl}^2)$ etc. are associated Laguerre polynomials. The radial solutions yields an independent quark bound-state condition in the form of a cubic equation:
\begin{equation}
\sqrt{(\omega_q/a)} (E_q^{\prime}-m_q^{\prime})=(4n+2l-1)
\end{equation}
The solution of the cubic equation provides the zeroth order binding energies of 
the confined quark and antiquark for all possible eigenmodes.

In the relativistic independent particle picture of this model, the constituent quark 
and antiquark are thought to move independently inside the $B_c$-meson bound state 
with momentum $\vec p_b$ and $\vec p_c$, respectively. Their individual momentum probability 
amplitudes are obtained in this model via momentum projection of respective quark orbitals (A1) in following forms:
For ground state mesons:($n=1$,$l=0$)
\begin{eqnarray}
G_b(\vec p_b)&=&{{i\pi {\cal N}_b}\over {2\alpha _b\omega _b}}
\sqrt {{(E_{p_b}+m_b)}\over {E_{p_b}}}(E_{p_b}+E_b)\exp {(-{
		{\vec p}^2\over {4\alpha_b}})}\nonumber\\
{\tilde G}_c(\vec p_c)&=&-{{i\pi {\cal N}_c}\over {2\alpha _c\omega _c}}
\sqrt {{(E_{p_c}+m_c)}\over {E_{p_c}}}(E_{p_c}+E_c)\exp {(-{
		{\vec p}^2\over {4\alpha_c}})}
\end{eqnarray}

For the excited meson state:(n=2, l=0)
\begin{eqnarray}
G_b(\vec p_b)&=&{{i\pi {\cal N}_b}\over {2\alpha _b}}
\sqrt {{(E_{p_b}+m_b)}\over {E_{p_b}}} {(E_{p_b}+E_b)\over{(E_b+m_b)}} 
({\vec {p_b}^2\over {2\alpha _b}}-{3\over 2})
\exp {(-{{\vec p_b}^2\over {4\alpha_b}})}\nonumber\\
{\tilde G}_c(\vec p_c)&=&{{i\pi {\cal N}_c}\over {2\alpha _c}}
\sqrt {{(E_{p_c}+m_c)}\over {E_{p_c}}}
{(E_{p_c}+E_c)\over {(E_c+m_c)}}
({\vec {p_c}^2\over {2\alpha _c}}-{3\over 2})
\exp {(-{
		{\vec p_c}^2\over {4\alpha_c}})}
\end{eqnarray}                   		
For the excited meson state (n=3, l=0)
\begin{eqnarray}  
G_b(\vec p_b)&=&{{i\pi {\cal N}_b}\over {2\alpha _b}}
\sqrt {{(E_{p_b}+m_b)}\over {E_{p_b}}} {(E_{p_b}+E_b)\over{(E_b+m_b)}} 
({\vec {p_b}^4\over {8 {\alpha _b}^2}}-{{5{\vec p_b}^2}\over {4\alpha_b}}+{{15\over 8}})
\exp {(-{{\vec p_b}^2\over {4\alpha_b}})}\nonumber\\
{\tilde G}_c(\vec p_c)&=&{{i\pi {\cal N}_c}\over {2\alpha _c}}
\sqrt {{(E_{p_c}+m_c)}\over {E_{p_c}}}
{(E_{p_c}+E_c)\over {(E_c+m_c)}}
({\vec {p_c}^4\over {8 {\alpha _c}^2}}-{{5{\vec p_c}^2}\over {4\alpha_c}}+{{15\over 8}})
\exp {(-{{\vec p_c}^2\over {4\alpha_c}})}
  \end{eqnarray} 
The binding energies of constituent quark and antiquark for the ground state of $B_c$ meson as well as the ground and excited final meson states for n=1,2,3; l=0 can also be obtained by solving respective cubic equations representing appropriate bound state conditions.

\begin{acknowledgments}
	The library and computational facilities provided by the authorities of Siksha 'O' Anusandhan Deemed to be University, Bhubanaeswar, 751030, India are duly acknowledged.
\end{acknowledgments}

\end{document}